\documentclass[ALICE,manyauthors]{cernphprep}

\usepackage{hyperref}
\usepackage{graphicx}  
\usepackage{dcolumn}   
\usepackage{bm}        
\usepackage{amssymb}   
\usepackage{amsmath}
\usepackage{amsfonts}
\usepackage{graphics}
\usepackage{grffile}   
\usepackage{epsfig}
\usepackage{units}
\usepackage[usenames]{color}
\usepackage[normalem]{ulem} 
\usepackage{color}
\usepackage[utf8]{inputenc}
\usepackage[T1]{fontenc}
\usepackage{subfigure}
\usepackage{lineno}
\usepackage{cite} 
\usepackage{orcidlink}

\usepackage{enumitem}
\setlist{itemsep=0pt, topsep=1.5mm, leftmargin=8mm}

\newcommand{\pp}{\ensuremath{\mathrm {p\kern-0.05em p}}}

\newcommand{\GeVc}{\ensuremath{\mathrm{GeV}\kern-0.05em/\kern-0.02em c}}

\newcommand{\pT}{\ensuremath{p_{\mathrm{T}}}}

\newcommand{\pTjet}{\ensuremath{p_{\mathrm{T}}^{\mathrm{jet}}}}
\newcommand{\pTchjet}{\ensuremath{p_{\mathrm{T}}^{\rm{ch\; jet}}}}
\newcommand{\pTtruth}{\ensuremath{p_{\mathrm{T,truth}}^{\mathrm{ch\; jet}}}}
\newcommand{\pTdet}{\ensuremath{p_{\mathrm{T,det}}^{\mathrm{ch\; jet}}}}
\newcommand{\pTsub}{\ensuremath{p_{\mathrm{T,subleading}}}}
\newcommand{\pTlead}{\ensuremath{p_{\mathrm{T,leading}}}}
\newcommand{\pTsubi}{\ensuremath{p_{\mathrm{T,subleading,i}}}}
\newcommand{\pTleadi}{\ensuremath{p_{\mathrm{T,leading,i}}}}
\newcommand{\tg}{\ensuremath{\theta_{\mathrm{g}}}}

\newcommand{\tgNP}{\ensuremath{\theta_{\mathrm{g}}^{\mathrm{NP}}}}
\newcommand{\Rg}{\ensuremath{R_{\mathrm{g}}}}
\newcommand{\zg}{\ensuremath{z_{\mathrm{g}}}}
\newcommand{\zgtruth}{\ensuremath{z_{\mathrm{g, truth}}}}
\newcommand{\zgdet}{\ensuremath{z_{\mathrm{g, det}}}}
\newcommand{\zcut}{\ensuremath{z_{\mathrm{cut}}}}

\newcommand{\etajet}{\ensuremath{\eta_{\mathrm{jet}}}}
\newcommand{\Ninc}{\ensuremath{N_{\mathrm{jet}}}}
\newcommand{\Sinc}{\ensuremath{\sigma_{\mathrm{jet}}}}

\begin{document}


\begin{titlepage}
\PHyear{2022}
\PHnumber{056}      
\PHdate{17 March}  
%

\title{Measurements of the groomed jet radius and momentum splitting fraction with the soft drop and dynamical grooming algorithms in pp collisions at $\bf{\sqrt{\textit{s}}=5.02}$ TeV}
\ShortTitle{Measurements of the groomed jet radius and groomed splitting fraction}   

\Collaboration{ALICE Collaboration\thanks{See Appendix~\ref{app:collab} for the list of collaboration members}}
\ShortAuthor{ALICE Collaboration} 

\date{\today}
\begin{abstract}

This article presents measurements of the groomed jet radius and
momentum splitting fraction in \pp{} collisions at {$\sqrt{s}=5.02$ TeV} with the ALICE detector at the Large Hadron Collider. 
Inclusive charged-particle jets are reconstructed at midrapidity using the anti-$k_{\rm{T}}$ algorithm
for transverse momentum $60<\pTchjet<80 \; \GeVc$.
We report results using two different grooming algorithms: soft drop and, for the first time, dynamical grooming.
For each grooming algorithm, a variety of grooming settings are used in order to explore the impact of collinear 
radiation on these jet substructure observables.
These results are compared to perturbative calculations that include resummation of large logarithms at all orders in the strong coupling constant.
We find good agreement of the theoretical predictions with the data for all grooming settings considered.

\end{abstract}

\end{titlepage}
\setcounter{page}{2}


\section{Introduction}

Measurements of high-energy jets produced in proton--proton collisions 
provide opportunities to test perturbative calculations 
and study non-perturbative (NP) effects in quantum chromodynamics (QCD)~\cite{Larkoski:2017jix,Asquith:2018igt,Marzani:2019hun}.
Jets also can be used to probe the properties of the quark--gluon plasma
by comparing jet observables in high-energy heavy-ion collisions to reference measurements in proton--proton collisions~\cite{Bjorken:1982qr, Adams:2005dq, Adcox:2004mh, Muller:2012zq, Braun-Munzinger:2015hba, Busza:2018rrf, Qin:2015srf, Blaizot:2015lma, Majumder:2010qh}.

Jet grooming techniques, such as soft drop~\cite{Larkoski:2014wba, Dasgupta:2013ihk, Larkoski:2015lea}
and dynamical grooming~\cite{Mehtar-Tani:2019rrk, Mehtar-Tani:2020oux, Caucal:2021bae, Caucal:2021cfb},
reduce the magnitude of non-perturbative contributions to jet substructure cross sections in \pp{} collisions by selectively removing soft 
large-angle radiation. This allows for well-controlled comparisons of measurements 
to perturbative QCD (pQCD) calculations.
Grooming techniques have also previously been applied to heavy-ion collisions,
in order to explore whether the quark--gluon plasma 
modifies the hard substructure of jets
\cite{PhysRevLett.119.112301, Mehtar-Tani:2016aco, Chang:2019nrx, Milhano:2017nzm, KunnawalkamElayavalli:2017hxo, Caucal:2019uvr, Ringer:2019rfk, Casalderrey-Solana:2019ubu, Andrews:2018jcm, Mulligan:2020tim, Caucal:2021cfb}.
Several measurements of groomed jet observables
have been made in \pp{} and heavy-ion collisions at the LHC and RHIC
~\cite{ALICE:2021njq, ALICE:2021obz, ATLAS:2019mgf, CMS:2017qlm, CMS:2018ypj, CMS:2018fof, STAR2020, STAR:2021kjt},
as well as in e$^+$e$^-$ collisions~\cite{Chen:2021uws}.
The benefits of different jet grooming algorithms remain a topic of ongoing study, since different grooming algorithms have different perturbative structure and
offer different flexibility via grooming parameters that can be adapted to
specific physics goals in either proton--proton or heavy-ion collisions (see e.g. Refs.~\cite{Mulligan:2020tim, Caucal:2021cfb, Ringer:2019rfk}).
In this article, we explore both the soft drop and dynamical grooming algorithms, and test the ability of pQCD calculations to describe
their behavior for a variety of grooming parameters.

\begin{figure}[!t]
\centering{}
\includegraphics[scale=0.39]{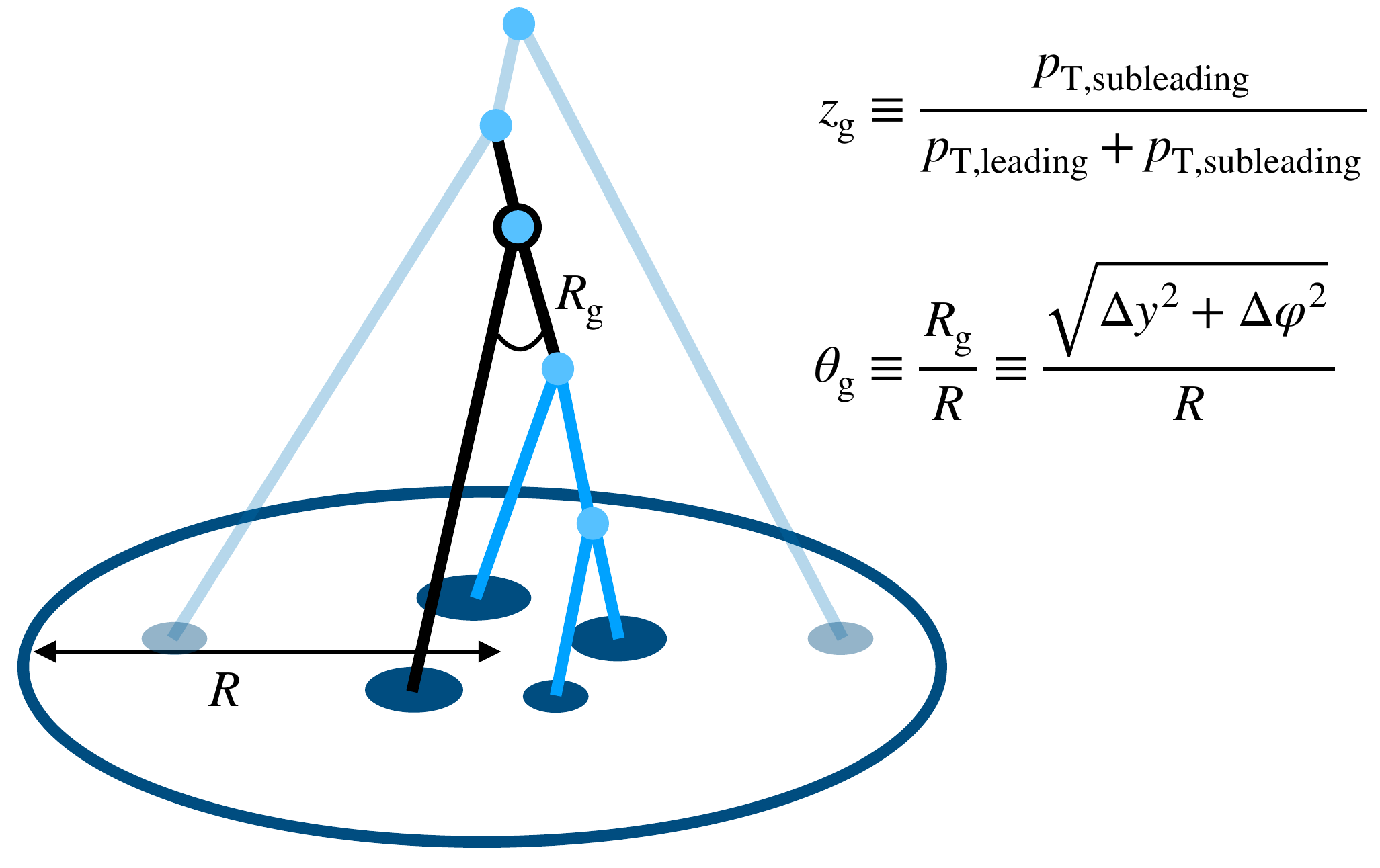}
\caption{Graphical representation of the angularly-ordered Cambridge--Aachen reclustering of jet constituents
and subsequent grooming procedure,
with the identified  splitting denoted in black and the splittings that were groomed away in light blue.
}
\label{fig:diagram}
\end{figure}

Jet grooming algorithms rely on procedures to recluster the constituents 
of reconstructed jets into a structure that better isolates perturbative
emissions in the jet.
One such structure is the primary Lund plane,
which approximately represents the angular and momentum 
phase space of partonic emissions off the leading hard-scattered parton.
The soft drop and dynamical grooming algorithms each identify a single splitting 
in the primary Lund plane~\cite{Dreyer:2018nbf} that satisfies a grooming condition.
The two algorithms are further described in Section~\ref{sec:analysis}.
In this article, we consider two observables that define the kinematics of the
identified splitting: 
\zg{}, the groomed jet momentum splitting fraction, and \tg{}, 
the (scaled) groomed jet radius, as shown in Fig.~\ref{fig:diagram}.
The groomed jet momentum splitting fraction is defined as the fraction of 
transverse momentum (\pT{}) relative to the beam that the sub-leading prong 
in the splitting carries relative to its parent:

\begin{equation} \label{eq:3}
\zg \equiv \frac{\pTsub} {\pTlead + \pTsub}.
\end{equation}

The (scaled) groomed jet radius is defined as the angular
distance between the two prongs of the identified hard splitting

\begin{equation} \label{eq:2}
\tg \equiv \frac{\Rg}{R} \equiv \frac{\sqrt{\Delta y ^2 + \Delta \varphi ^2}}{R},
\end{equation}

where $R$ is the jet radius
and \Rg{} is the rapidity--azimuth ($y$--$\varphi$) separation of the identified splitting.

The soft drop \zg{} and \tg{} distributions
have recently been calculated in \pp{} collisions at Next-to-Leading Logarithmic (NLL$^\prime$)
accuracy~\cite{Kang:2019prh, Cal:2021fla}. 
Measurements of \zg{} and \tg{} serve to test these analytical predictions, 
in particular, the role of beyond-LL pQCD effects, as well as constrain
the role of non-perturbative effects.
Moreover, by measuring these observables for a variety of grooming conditions $\beta$ (see Section~\ref{sec:grooming} for further details),
one can systematically study the role of collinear radiation in jet substructure,
since increasing $\beta$ removes less and less collinear radiation in the grooming process.
Measurements of both \zg{} and \tg{} for $\beta=0,1$, and 2 
have been performed by the ATLAS Collaboration~\cite{ATLAS:2019mgf} for dijet events with leading $\pTjet>300\;\GeVc$,
and several measurements of \zg{} and \tg{} have been performed for $\beta=0$
across a wide range of jet \pT{}~\cite{ALICE:2021obz, CMS:2017qlm, STAR2020, Chen:2021uws}.
In this article, we complement these studies by 
measuring \zg{} and \tg{} for $\beta=0,1$, and 2 for $60<\pTchjet<80 \; \GeVc$.

The dynamically groomed \zg{} and \tg{} distributions
have recently been calculated in \pp{} collisions
at Next-to-Next-to-Double Logarithm (N$^2$DL)
accuracy~\cite{Mehtar-Tani:2019rrk, Caucal:2021bae}.
In this article, we perform the first measurement of dynamically groomed jet substructure observables,
providing the first test of these calculations.

We report measurements in \pp{} collisions at center-of-mass collision energy {$\sqrt{s}=5.02$ TeV} with the ALICE detector.
Charged-particle jets are reconstructed in the pseudorapidity range $|\etajet|<0.5$ for jet radius $R=0.4$ with $60 < \pTchjet{} < 80 \; \GeVc$.
The \zg{} and \tg{} distributions are measured using both the soft drop and dynamical grooming procedures,
each with a variety of grooming settings.
These results are compared to pQCD calculations as well as the PYTHIA8~\cite{Sjostrand:2014zea,Skands:2014pea} Monte Carlo (MC) event generator. 
While track-based jet observables are collinear-unsafe~\cite{Chang:2013rca, Chen:2020vvp, Chien:2020hzh},
they can be measured with greater precision than calorimeter-based jet observables,
and recent measurements have demonstrated that for many substructure observables 
track-based distributions are compatible with the corresponding collinear-safe distributions~\cite{ATLAS:2019mgf}.
Comparisons of theoretical calculations to our track-based jet substructure measurements are discussed further in Section~\ref{sec:results}.

\section{Experimental setup and data sets} \label{sec:setup}
A description of the ALICE detector and its performance can be found in
Refs.~\cite{aliceDetector, ALICE:2014sbx}. The \pp{} data set used in this
analysis was collected in 2017 during LHC Run 2 at $\sqrt{s} = 5.02$ TeV 
using a minimum-bias trigger defined by the coincidence of the signals from two
scintillator arrays in the forward region (V0 detectors)~\cite{ALICE:2013axi}. 
The event selection includes a primary vertex selection, where
the primary vertex is required to be within 10 cm from the center of the detector along the beam direction. 
Events with more than one reconstructed primary vertex were classified as pileup and rejected~\cite{Acharya:2019jyg}.
After these selections, the pp data sample
contains 870 million events and corresponds to an integrated luminosity of
$18.0 \pm 0.4$ nb$^{-1}$~\cite{ppXsec}.

The analysis uses charged-particle tracks reconstructed with information from the Time Projection Chamber
(TPC)~\cite{Alme_2010} and the Inner Tracking System (ITS)~\cite{ALICE:2010tia}.
Two types of tracks are defined: global tracks and complementary tracks.
Global tracks are required to include at least
one hit in the silicon pixel detector (SPD) comprising the first
two layers of the ITS and to satisfy several track quality selections.
Complementary tracks are all those satisfying all the selection
criteria of global tracks except for the request of a point in the SPD. They are refitted using
the primary vertex to constrain their trajectory in order to preserve a good momentum
resolution, especially at high transverse momentum.
Including this second class of tracks ensures approximately uniform azimuthal acceptance, 
while preserving similar \pT{} resolution to tracks with SPD hits. 
Tracks with $0.15 < \pT < 100 \;\GeVc$ are 
accepted over pseudorapidity range $|\eta| < 0.9$ and azimuthal angle $0 < \varphi < 2\pi$. 

The instrumental performance of the detector is estimated 
with a MC simulation done using PYTHIA8~\cite{Sjostrand:2014zea} with the Monash 2013 tune ~\cite{Skands:2014pea}
 for the event generation and GEANT3~\cite{Brun:1119728} for the transport code propagating particles through the simulated ALICE apparatus. 
The tracking efficiency in \pp{} collisions is approximately 67\% 
at track $\pT=0.15 \;\GeVc$, and rises to approximately 84\% at $\pT=1 \;\GeVc$, and remains above 75\% at higher \pT.
The momentum resolution $\sigma(\pT)/\pT$ was estimated from the covariance matrix of the track fit~\cite{ALICE:2014sbx},
and is approximately 1\% at track $\pT=1$ \GeVc{} and 4\% at ${\pT=50 \;\GeVc}$.

\section{Analysis method}
\label{sec:analysis}

Jets are reconstructed from charged-particle tracks with FastJet 3.2.1~\cite{Cacciari:2011ma} using the
anti-$k_{\mathrm{T}}$ algorithm with $E$-scheme recombination
with resolution parameter $R=0.4$~\cite{Cacciari:2008gp, Cacciari:2008gn}.
All tracks are assigned a mass equal to the $\pi^{\pm}$ meson mass.
The jet axis is required to be within the fiducial volume of the TPC, $\left| \etajet \right| < 0.5$, where $\etajet$ is the jet pseudorapidity.
The jet reconstruction performance for this data set is described in Ref.~\cite{ALICE:2021njq}.
The underlying event (UE) consists of approximately $\pT=1\;\GeVc$ per jet, and is not subtracted. Therefore, UE corrections must be included in theoretical calculations when comparing to the data.

\subsection{Grooming algorithms}
\label{sec:grooming}

The soft drop and dynamical grooming algorithms identify a single splitting 
in the primary Lund plane~\cite{Dreyer:2018nbf} that satisfies a grooming condition.
The $i^{\rm{th}}$ splitting in the primary Lund plane is defined by
\begin{equation} \label{eq:4}
\begin{split}
z_i &\equiv \frac{\pTsubi} {\pTleadi + \pTsubi}, \\
\theta_i &\equiv \frac{\Delta R_i}{R}, 
\end{split}
\end{equation}

where $\Delta R_i = \sqrt{\Delta y_i ^2 + \Delta \varphi_i ^2}$ is the rapidity-azimuth separation of the $i^{\rm{th}}$ splitting.
Note that when reconstructing the primary Lund plane, one must choose a reclustering radius $R_{\rm{recluster}}$; for soft drop $R_{\rm{recluster}}=R$ is used, which results in $\tg\leq1$,
whereas for our implementation of dynamical grooming $R_{\rm{recluster}}=\infty$ is used, which results in $\tg>1$ for <1\% of cases (which we neglect).

In the soft drop grooming algorithm, the grooming condition is given by
\begin{equation} \label{eq:5}
z_i > \zcut \theta_i^\beta,
\end{equation}
where $\zcut$ and the exponent $\beta$ are tunable free parameters of the grooming algorithm.
The first such splitting to pass the grooming condition defines the soft drop groomed jet splitting.
As the grooming parameter $\beta$ increases, the quantity $\zcut \theta_i^\beta$ 
becomes small for collinear radiation. This causes the algorithm to be  less likely to drop collinear radiation ---
corresponding to less grooming overall, and particularly less grooming for collinear radiation.
Note that for the values $\beta\geq0$ considered here, 
\zg{} is Sudakov safe~\cite{Larkoski:2015lea}
and \tg{} is infrared-collinear safe~\cite{Kang:2019prh}.

The dynamical grooming algorithm, on the other hand, identifies the splitting that maximizes
\begin{equation} \label{eq:6}
z_i (1-z_i)p_{\rm{T},i} \theta_i^a
\end{equation}
over all splittings in the primary Lund plane,
where the exponent $a$ is a continuous free parameter. 
The grooming parameter $a$ defines the density with which the phase space
of the Lund plane is groomed away.
The case $a\rightarrow 0$ selects the splitting
with largest $z$, and is somewhat similar to soft drop with $\beta=0$, 
which grooms away splittings below a certain $z$.
The case $a=1$ selects the splitting with largest transverse momentum, and is roughly analogous to soft drop with $\beta=-1$, which grooms away splittings below a certain transverse momentum (see Ref.~\cite{Mehtar-Tani:2019rrk} for further details). 
Since the grooming condition in dynamical grooming defines a maximum rather than an explicit cut 
(as in the case of soft drop), every dynamically groomed jet will always return a splitting,
whereas in soft drop it is possible that a jet does not contain any splitting satisfying the grooming condition.

\subsection{Corrections}
The reconstructed \pTchjet{} and \zg{} (\tg{}) differ from
their true values due to tracking inefficiency,
particle--material interactions, and track \pT{} resolution. 
To account for these effects, events are simulated
using PYTHIA8 Monash 2013~\cite{Sjostrand:2014zea,Skands:2014pea} for the event generation and GEANT3~\cite{Brun:1119728} for the transport code propagating particles through the simulated ALICE apparatus,
as described in Section~\ref{sec:setup}.
The truth-level jets are constructed from the charged primary
particles of the PYTHIA8 event, defined as all particles with a 
mean proper lifetime larger than 1 cm/$c$, and excluding the decay
products of these particles~\cite{primaryParticleALICE}.
A 4D response matrix is constructed that describes
the detector response in \pTchjet{} and \zg{} (and similarly for \tg{}): 
$R\left( \pTdet, \pTtruth, \zgdet, \zgtruth \right)$, where \pTdet{} is the detector-level \pTchjet{} 
and \pTtruth{} is the truth-level \pTchjet{}.

Then, a 2D unfolding is performed in \pTchjet{} and \zg{} using the iterative Bayesian
unfolding algorithm~\cite{DAgostini, dagostini2010improved} implemented in the RooUnfold package~\cite{roounfold}.
The distributions are corrected for ``misses'', in which a jet exists inside the considered truth level range but not inside the detector level range.
The rate of ``fakes'', in which a jet exists inside the considered detector level range but not inside the truth level range,
is negligible.
The number of iterations, which sets the strength of regularization, is
chosen by minimizing the quadrature 
sum of the statistical and systematic unfolding uncertainties.
This results in the optimal number of iterations equal to 3 in all cases. 

To validate the performance of the unfolding procedure,
refolding tests are performed, in which the response matrix is multiplied by the unfolded solution
and compared to the original detector-level spectrum. Closure tests are also performed,
in which the shape of the generated MC spectrum
is modified to account for the fact that the true distribution may
be different from the MC spectrum. In all cases,
successful closure within statistical and systematic uncertainties is achieved.

\section{Systematic uncertainties}

Systematic uncertainties due to the tracking
efficiency, the unfolding procedure, and the MC generator model dependence are considered.
Table~\ref{tab:sysunc} summarizes the systematic uncertainty contributions from each of these sources.
The total systematic uncertainty is calculated as the sum in quadrature of all of the individual systematic uncertainties described below.

\begin{table}[!t]
\centering
\caption{Summary of systematic uncertainties on unfolded \zg{} and \tg{} distributions for $60<\pTchjet<80\;\GeVc$.
The ranges correspond to the minimum and maximum systematic uncertainties obtained.}
\begin{tabular}{ l ccccc }
\tabularnewline \hline \hline & \multicolumn{4}{c}{Relative uncertainty (\%)}
\tabularnewline \hline Soft drop, $\zcut=0.1$ & Tracking efficiency & Unfolding & Generator & Total \\
\hline
\zg \qquad $\beta = 0$ & 0--2\% & 0--4\% & 0--1\% & 2--5\% \\
\;\;\;\qquad $\beta = 1$ & 0--4\% & 0--4\% & 0--3\% & 1--6\% \\
\;\;\;\qquad $\beta = 2$ & 0--3\% & 1--5\% & 0--5\% & 2--7\% \\
\tg \qquad $\beta = 0$ & 2--8\% & 2--6\% & 0--4\% & 3--9\% \\
\;\;\;\qquad $\beta = 1$ & 2--10\% & 0--5\% & 0--3\% & 2--12\% \\
\;\;\;\qquad $\beta = 2$ & 2--11\% & 1--6\% & 0--5\% & 2--13\% \\
\hline Dynamical grooming & Tracking efficiency & Unfolding & Generator & Total \\
\hline
\zg \qquad $a = 0.1$ & 0--14\% & 2--10\% & 0--4\% & 2--17\% \\
\;\;\;\qquad$a = 1.0$ & 0--5\% & 1--4\% & 0--2\% & 1--5\% \\
\;\;\;\qquad$a = 2.0$ & 0--4\% & 1--5\% & 0--3\% & 2--7\% \\
\tg \qquad $a = 0.1$ & 0--6\% & 1--5\% & 0--4\% & 2--8\% \\
\;\;\;\qquad$a = 1.0$ & 1--9\% & 1--5\% & 0--3\% & 2--10\% \\
\;\;\;\qquad$a = 2.0$ & 0--8\% & 1--3\% & 0--7\% & 2--11\% \\
\hline \hline

\end{tabular} 
\label{tab:sysunc}
\end{table}

The systematic uncertainty due to the uncertainty in the tracking efficiency is evaluated using random rejection of additional tracks in the jet finding. 
The tracking efficiency uncertainty, estimated from the variation of the track selection criteria and a detailed study of the ITS--TPC track-matching efficiency uncertainty, is 4\%.
In order to assign a systematic uncertainty to the nominal
result, an alternative response matrix is constructed by randomly rejecting an additional 4\% of
tracks in jet finding, and the unfolding procedure is repeated. 
This result is compared to the nominal result, with
the differences in each bin taken as the systematic uncertainty. 
The uncertainty on the track momentum resolution is a sub-leading
effect to the tracking efficiency and is taken to be negligible.

Four sets of variations of the unfolding procedure are performed
in order to estimate the systematic
uncertainty arising from the unfolding regularization procedure.
\begin{itemize}
    \item The number of iterations in the unfolding procedure is varied by $\pm2$ units, 
    chosen based on studies of the rate of convergence of the unfolding procedure. 
    The average difference with respect to the nominal result is taken as the systematic uncertainty.
    \item The prior distribution is scaled by a power law in \pTchjet{} 
    and by
	$\pT^{\pm0.5} \zg^{\pm0.5}$ for the \zg{} analysis. 
	For the \tg{} analysis, a linear scaling in \tg{} by $\pm 50\%$ over its reported range, scaling by $\pT^{\pm0.5} [1\pm0.5(2\tg-1)]$, is applied.
    The average difference between the result unfolded with this prior and the original 
    is taken as the systematic uncertainty. 
    \item The binnings in \zg{} and \tg{} are varied to be finer and coarser than the nominal binning.
    \item The lower bound in the detector level charged-particle jet transverse momentum \pTdet{} range is extended up and down by 5 \GeVc. 
\end{itemize}
The total unfolding systematic uncertainty is then the standard deviation of the variations, $\sqrt{\sum_{i=1}^{N} \sigma_{i}^{2} / N}$, 
where $N=4$ and $\sigma_{i}$ is the systematic uncertainty due to a single group of variations, since they each comprise independent estimates of the same underlying systematic uncertainty in the regularization.

The systematic uncertainty due to the model dependence of the generator used to construct the 
response matrix is estimated by comparing results obtained with PYTHIA8 Monash 2013~\cite{Sjostrand:2014zea,Skands:2014pea}
to that obtained with Herwig7 (default tune)~\cite{Bellm:2015jjp}.
The tracking efficiency and track \pT{} resolution are parameterized using
fast simulations and response matrices are built using these two generators.
These response matrices are then used to unfold the measured
data, and the differences between the two unfolded results in each
interval are taken as a symmetric uncertainty.

\section{Results}
\label{sec:results}

We report the \zg{} and \tg{} distributions in the \pTchjet{} interval between 60 and 80 GeV/$c$.
All presented results use $R=0.4$ jets reconstructed from charged particles at midrapidity,
and are corrected for detector effects.
The distributions are reported as normalized differential cross sections,
\begin{equation} \label{eq:7}
\frac{1}{\Sinc} \frac{\mathrm{d}\sigma}{\mathrm{d}\zg}
=
\frac{1}{\Ninc} \frac{\mathrm{d}N} {\mathrm{d}\zg},
\end{equation}
where \Ninc{} (\Sinc{}) is the number (cross section) of inclusive charged-particle jets within the given \pTchjet{} interval, and $N$ ($\sigma$) is the number (cross section) of groomed splittings.
The same normalization as in Eq.~\ref{eq:7} is used for \tg{}.
With this normalization, the integral of Eq.~\ref{eq:7}
is equal to the fraction of jets that pass the grooming condition.

\subsection{Soft drop}

\begin{figure}[!th]
\centering{}
\includegraphics[scale=0.48]{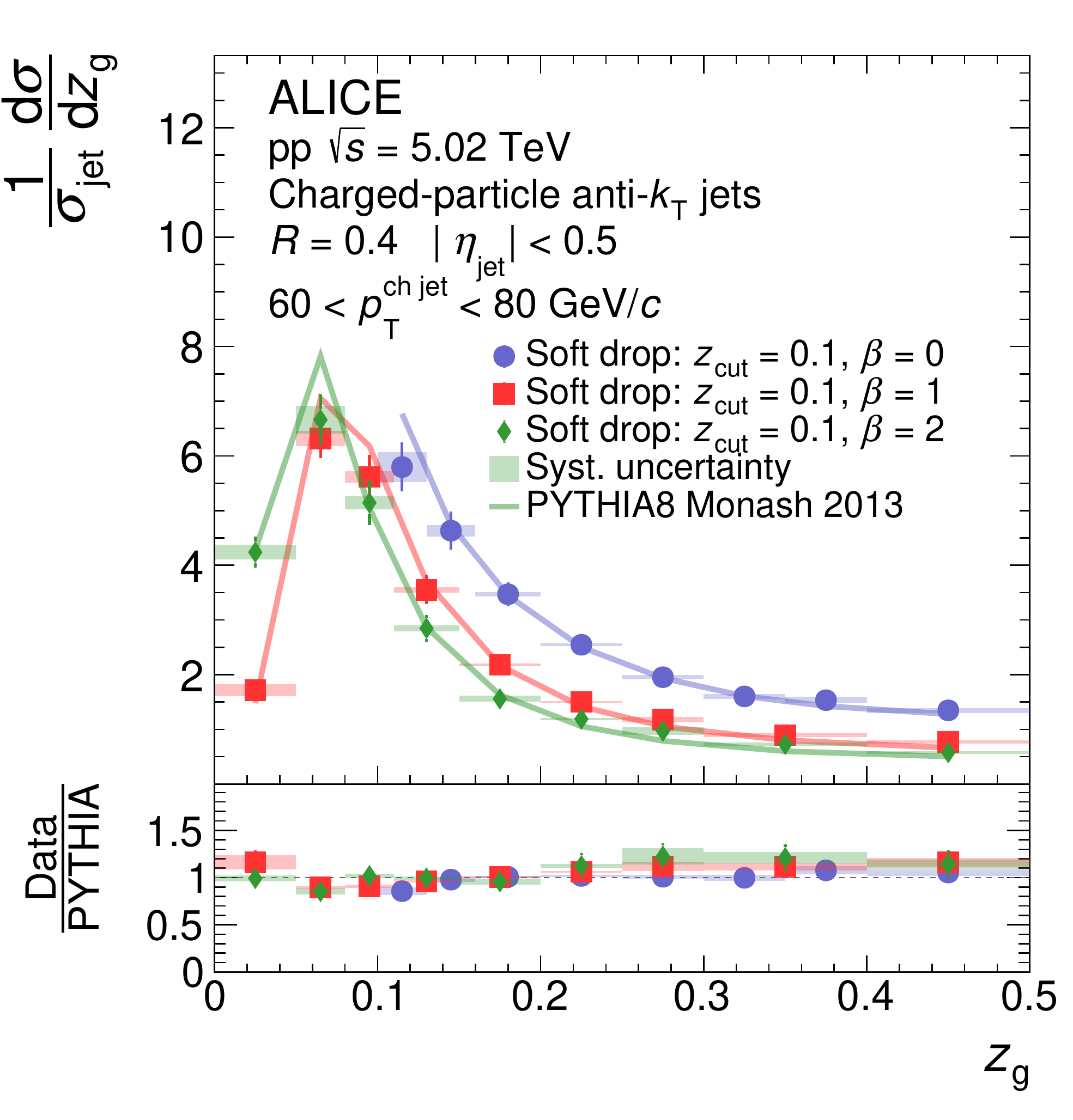}
\caption{ALICE measurements of \zg{} distributions 
in \pp{} collisions at $\sqrt{s}=5.02\;$ TeV with soft drop for three values of the grooming parameter $\beta$, compared with PYTHIA8 Monash 2013~\cite{Sjostrand:2014zea,Skands:2014pea} calculations.}
\label{fig:sd-zg}
\end{figure}

\begin{figure}[!bh]
\centering{}
\includegraphics[scale=0.48]{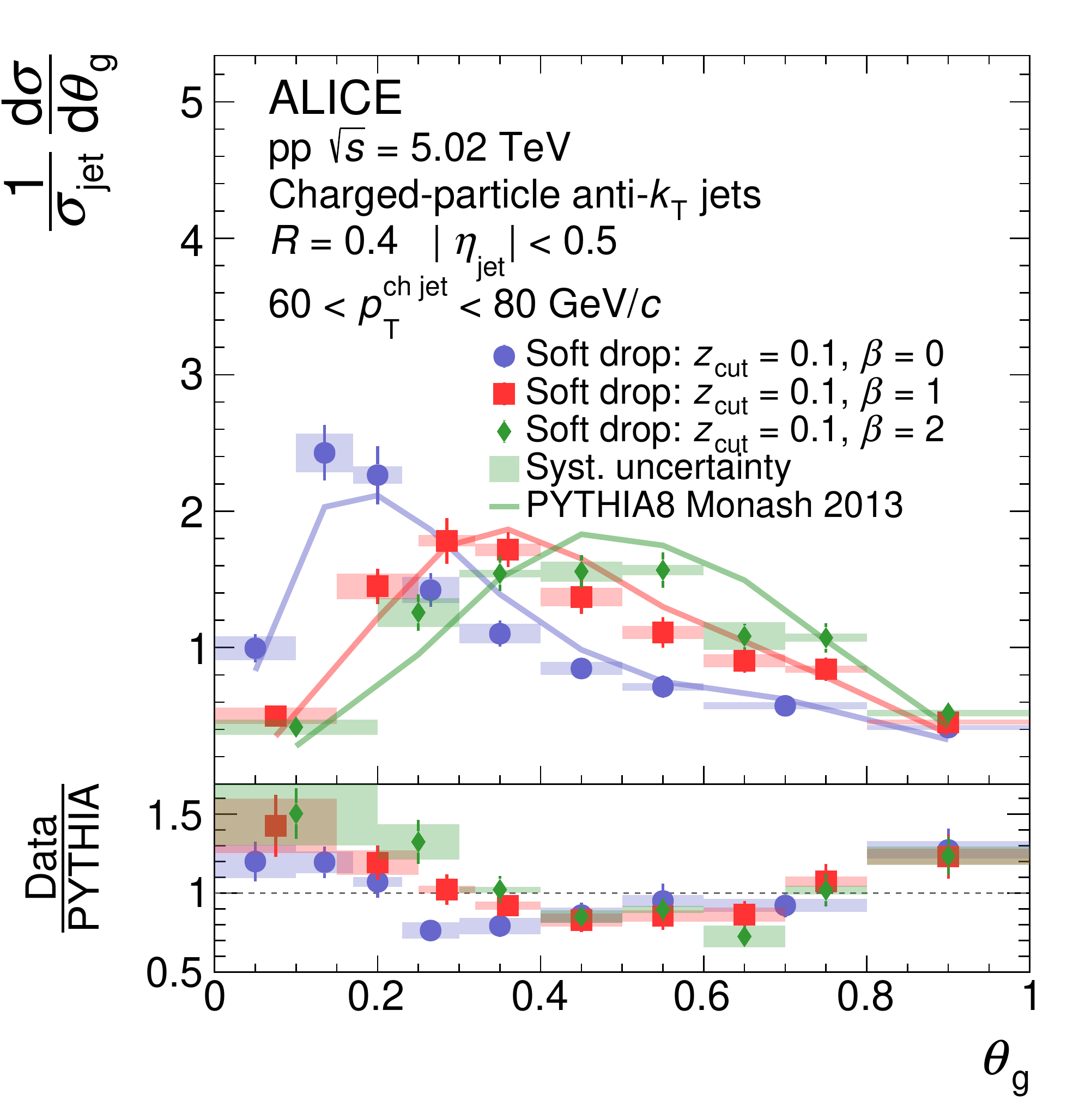}
\caption{ALICE measurements of \tg{} distributions 
in \pp{} collisions at $\sqrt{s}=5.02\;$ TeV with soft drop for three values of the grooming parameter $\beta$, compared with PYTHIA8 Monash 2013~\cite{Sjostrand:2014zea,Skands:2014pea} calculations.}
\label{fig:sd-tg}
\end{figure}

Figures~\ref{fig:sd-zg} and~\ref{fig:sd-tg} show the measured \zg{} and \tg{}
distributions for jets with soft drop grooming
for grooming parameters $\zcut=0.1$ and $\beta=0,1$, and 2.
The \zg{} distributions fall with increasing \zg{}, as is typical of
the Altarelli--Parisi splitting functions~\cite{Altarelli:1977zs}.
The \zg{} distribution for $\beta=0$ cannot populate $\zg<0.1$ due to the grooming condition.
However, for $\beta>0$ it is possible for a sufficiently narrow splittings 
with $\zg<\zcut$ to pass the grooming condition.
The \zg{} distributions are generally described by PYTHIA8~\cite{Sjostrand:2014zea} within approximately 20\%.
The \tg{} distributions exhibit a peak at increasingly large \tg{} as $\beta$ increases,
due to the angular component in the grooming condition.
The \tg{} distributions are described by PYTHIA8~\cite{Sjostrand:2014zea} typically within
20\% but with deviations at low \tg{} up to approximately 50\%.
Due to ill-defined perturbative accuracy in general-purpose MC generators
such as PYTHIA
and the fact that they are highly tuned to reproduce data, including jet-related observables~\cite{Skands:2014pea}, it is difficult to draw detailed
physics conclusions from their comparison to data. Because of this, we instead
turn our attention to comparisons with analytical calculations based on pQCD, where deeper insight can be obtained.

Theoretical calculations with soft drop grooming have been carried out
within the Soft-Collinear Effective Theory (SCET) framework~\cite{SCETplaceholder}
for \tg{}~\cite{Kang:2019prh} and \zg{}~\cite{Cal:2021fla}.
These calculations include all-order resummations of large logarithms to
NLL$^\prime$ accuracy~\cite{Kang:2019prh}. 
In order to compare these parton-jet predictions to our measurement using
charged-particle jets, a ``forward folding'' procedure is applied to account
for hadronization and charged-particle effects, followed by a bin-by-bin scaling
to account for Multi-Parton Interactions. These corrections are carried
out following the procedure outlined in Ref.~\cite{ALICE:2021njq}.
Given that the scale $\tg \zg \pT R$ becomes non-perturbative at low \tg{}, 
and that our measurements of the \zg{} distribution do not include a lower cutoff
in \tg{}, we forgo these comparisons for the \zg{} distribution and refer 
the reader to Ref.~\cite{Cal:2021fla}. 
Instead, we focus on comparison of the measured \tg{} distribution to the SCET calculations.

\begin{figure}[!t]
\centering
\hspace*{-1cm}\includegraphics[scale=0.85]{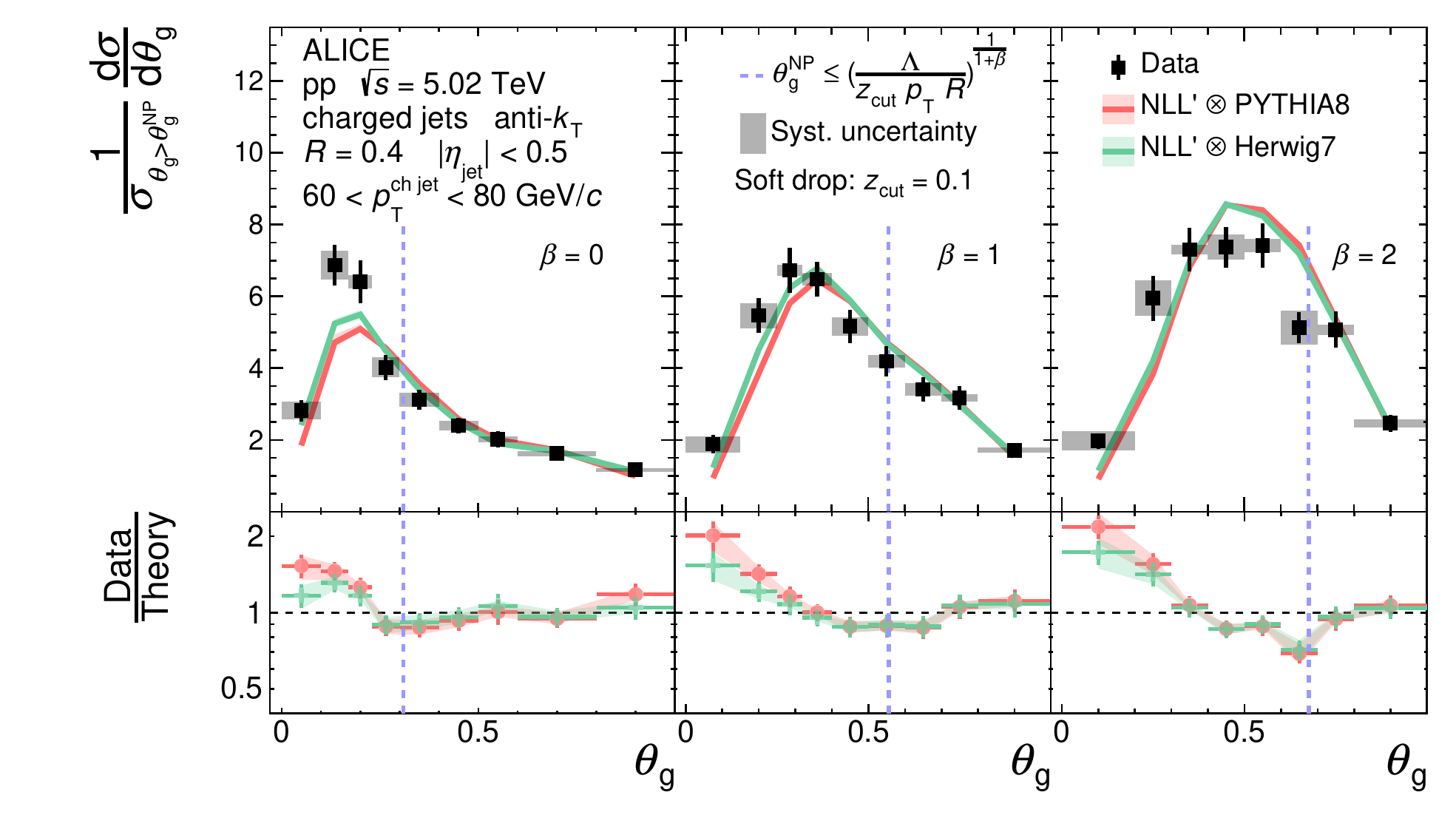}\vspace{-0.5cm}
\caption{ALICE measurements of \tg{} distributions in \pp{} collisions at $\sqrt{s}=5.02$ TeV with
soft drop, compared with NLL$^\prime$ predictions carried out with SCET ~\cite{Kang:2019prh} and
corrected for non-perturbative effects using either PYTHIA8~\cite{Sjostrand:2014zea}
or Herwig7~\cite{Bellm:2015jjp}. The distributions are normalized such that the
integral of the perturbative region defined by $\tg{} > \tgNP{}$
(to the right of the dashed vertical blue line) is unity.
The non-perturbative scale in Eq.~\ref{eq:8} is taken to be $\Lambda=1\;\GeVc$.
In determining the normalization, intervals that overlap with the dashed blue line are considered to be in the non-perturbative (left) region.}
\label{fig:tg-theory}
\end{figure}

Figure~\ref{fig:tg-theory} compares the measured \tg{} distributions with pQCD
calculations based on SCET~\cite{Kang:2019prh} using
either PYTHIA8~\cite{Sjostrand:2014zea} or Herwig7~\cite{Bellm:2015jjp} to account
for non-perturbative corrections.
The PYTHIA8 and Herwig7 corrections show generally similar behavior.
Systematic uncertainties on the analytical predictions are estimated by
systematically varying combinations of scales that emerge in the calculation.
The softest of these scales determines a transition between the 
perturbative and non-perturbative regimes:
\begin{equation} \label{eq:8}
\tgNP \lesssim \left( \frac{\Lambda}{\zcut \pT R} \right)^{\frac{1}{1+\beta}},
\end{equation}
where $\Lambda$ is the energy scale at which $\alpha_s$ becomes non-perturbative.
This transition is indicated by a dashed vertical blue line 
at $\Lambda=1\;\GeVc$, taking \pT{} to be the weighted average \pTchjet{} in
the considered interval scaled by 20\% to approximately translate the \pT{}
scale from charged-particle jets to full jets.
The cross section is normalized according to the integral of the distribution
in the perturbative region,

\begin{equation} \label{eq:9}
\frac{1}{\sigma_{\tg>\tgNP}} \frac{\mathrm{d}\sigma}{\mathrm{d}\tg},
\qquad\mathrm{where}\qquad 
\sigma_{\tg>\tgNP} = \int_{\tgNP}^{1} \frac{\mathrm{d}\sigma}{\mathrm{d}\tg}\mathrm{d}\tg.
\end{equation}

The measured \tg{} distributions agree with the SCET
calculations within uncertainties in the perturbative region (i.e. to the
right of the dashed line), whereas divergence is seen at low values of \tg{},
where non-perturbative effects dominate and the perturbative calculation
is expected to break down. This holds for all values of $\beta$. 
Note that the perturbative regime contains an increasingly small fraction
of the distribution as $\beta$ grows, which demonstrates that at these
\pTchjet{} values, the majority of the \tg{} distribution can only be captured by pQCD
for sufficiently small $\beta$.

\subsection{Dynamical grooming}

\begin{figure}[!h]
\centering{}
\includegraphics[scale=0.48]{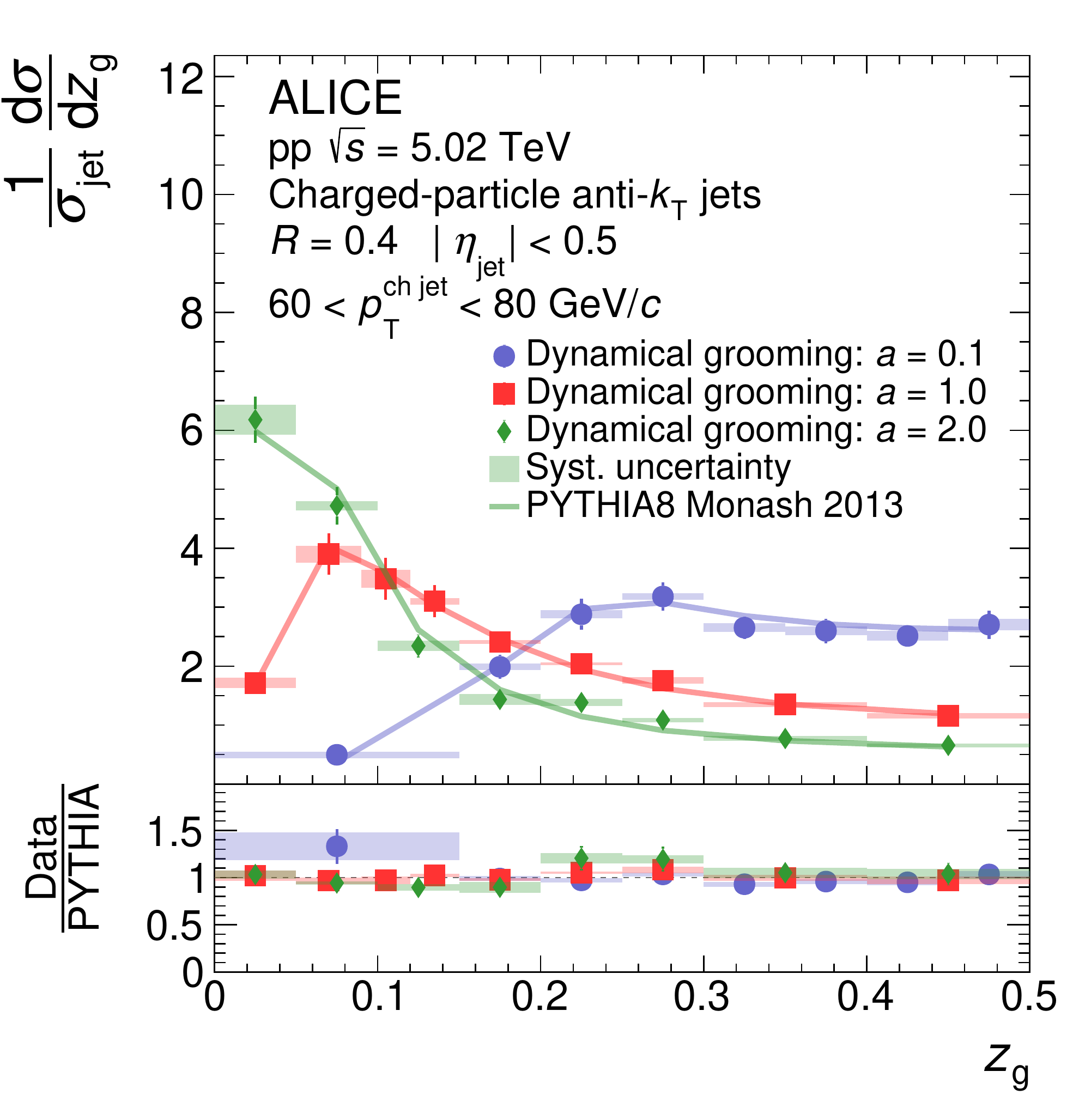}
\caption{ALICE measurements of \zg{} distributions 
in \pp{} collisions  at $\sqrt{s}=5.02\;$ TeV with dynamical grooming~\cite{Mehtar-Tani:2019rrk} for three values of the
grooming parameter $a$, compared with PYTHIA8 Monash 2013~\cite{Sjostrand:2014zea,Skands:2014pea} calculations.}
\label{fig:dg-zg}
\end{figure}

\begin{figure}[!h]
\centering{}
\includegraphics[scale=0.48]{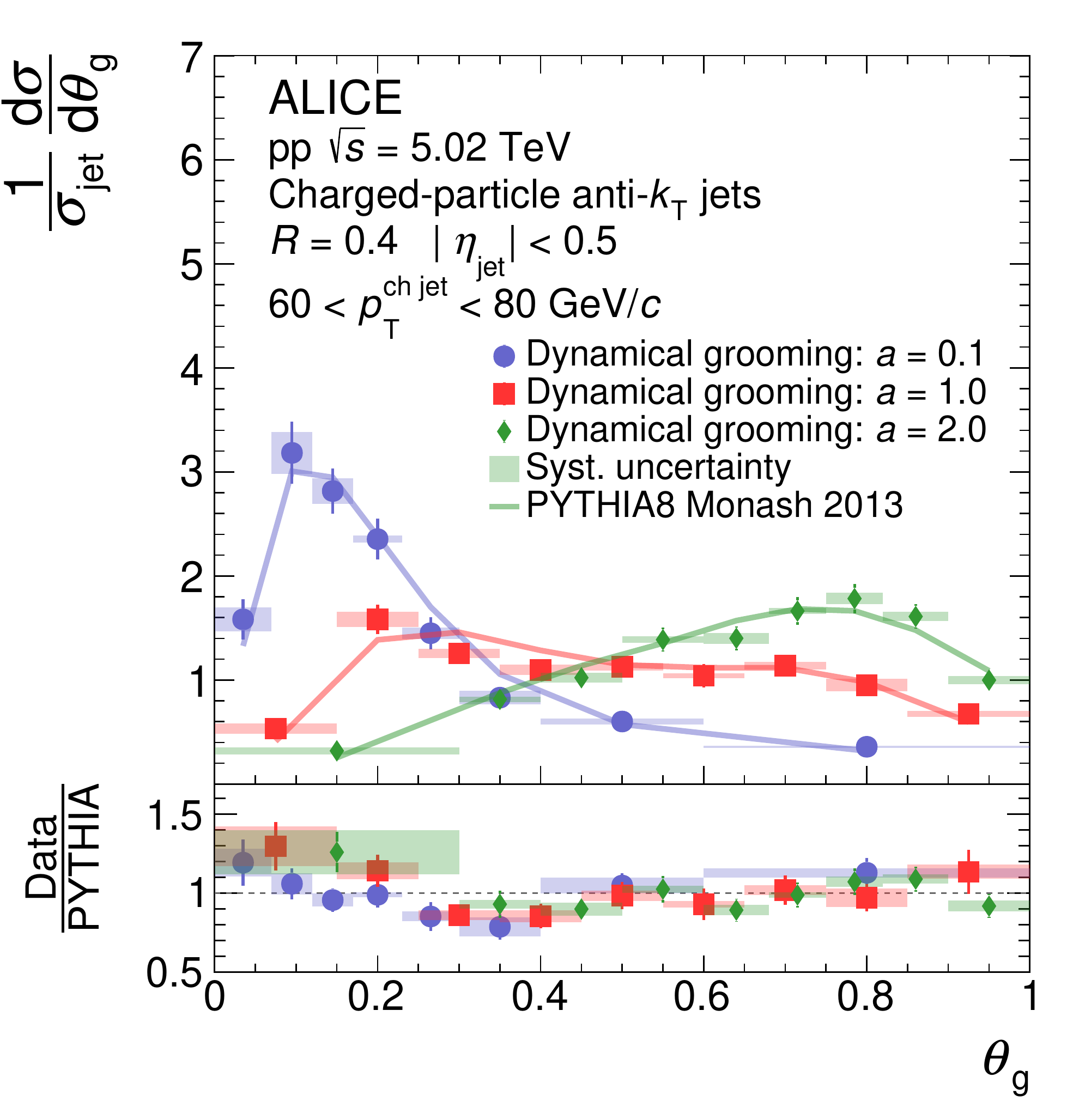}
\caption{ALICE measurements of \tg{} distributions 
in \pp{} collisions  at $\sqrt{s}=5.02\;$ TeV with dynamical grooming~\cite{Mehtar-Tani:2019rrk} for three values of the
grooming parameter $a$, compared with PYTHIA8 Monash 2013~\cite{Sjostrand:2014zea,Skands:2014pea} calculations.}
\label{fig:dg-tg}
\end{figure}

Figures~\ref{fig:dg-zg} and~\ref{fig:dg-tg} show the \zg{} and \tg{} distributions in \pp{} collisions
for jets with dynamical grooming
for several values of the grooming parameter $a$. 
For small values of $a$, the grooming condition favors splittings with symmetric
longitudinal momenta, which is reflected in the distributions skewing towards
large \zg{} and small \tg. As $a$ increases, the grooming condition favors 
splittings with large angular separation, which is reflected in the distributions 
skewing towards small \zg{} and large \tg.
The results are compared with PYTHIA8 Monash 2013~\cite{Sjostrand:2014zea,Skands:2014pea},
which generally describes the data within approximately 20\%.

\begin{figure}[!t]
\centering{}
\includegraphics[scale=0.8]{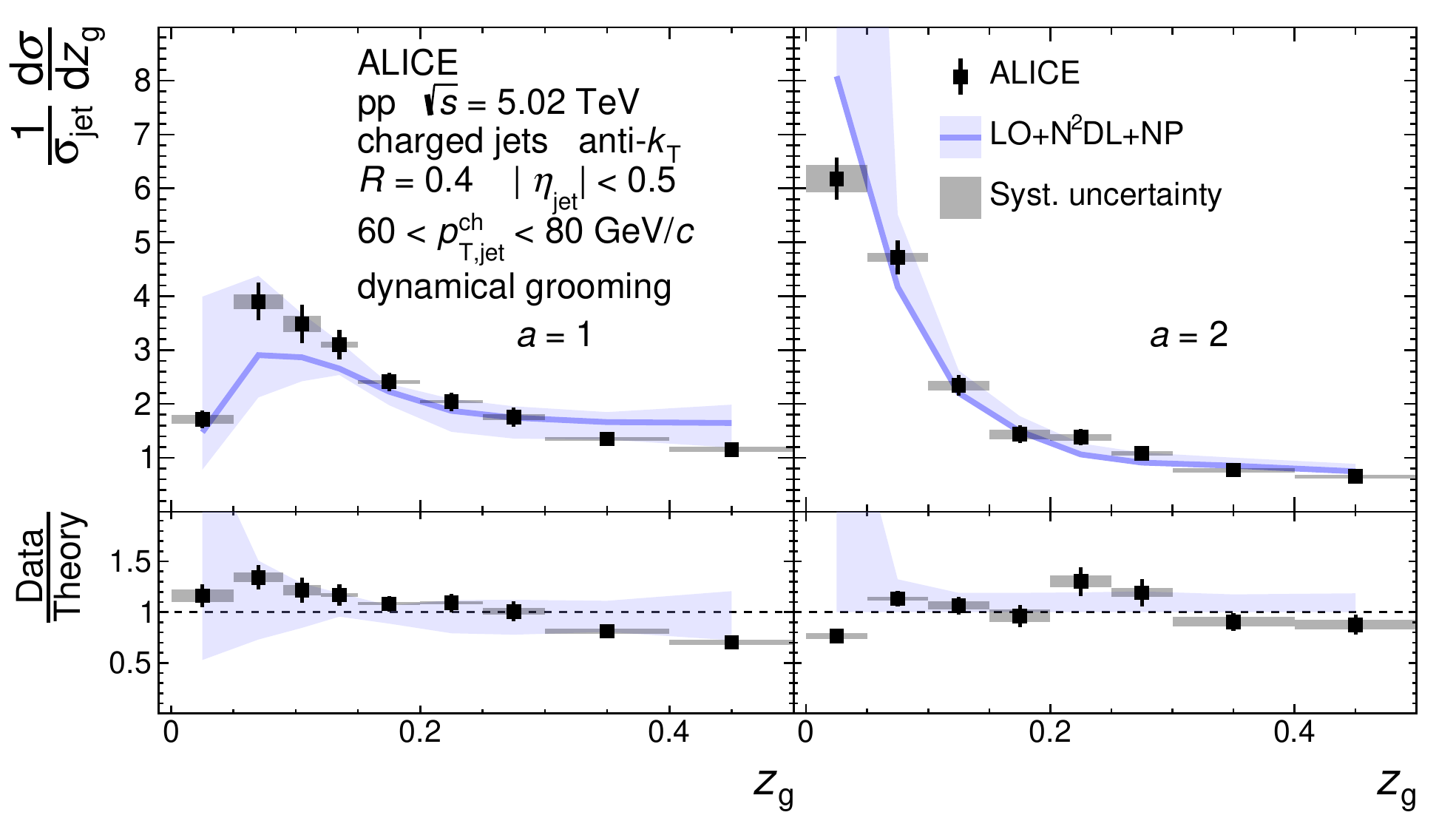}
\caption{ALICE measurements of \zg{} distributions 
in \pp{} collisions  at $\sqrt{s}=5.02\;$ TeV with dynamical grooming
for two values of the
grooming parameter $a$, compared with pQCD calculations~\cite{Mehtar-Tani:2019rrk, Caucal:2021bae}.}
\label{fig:dg-zg-theory}
\end{figure}

\begin{figure}[!t]
\centering{}
\includegraphics[scale=0.8]{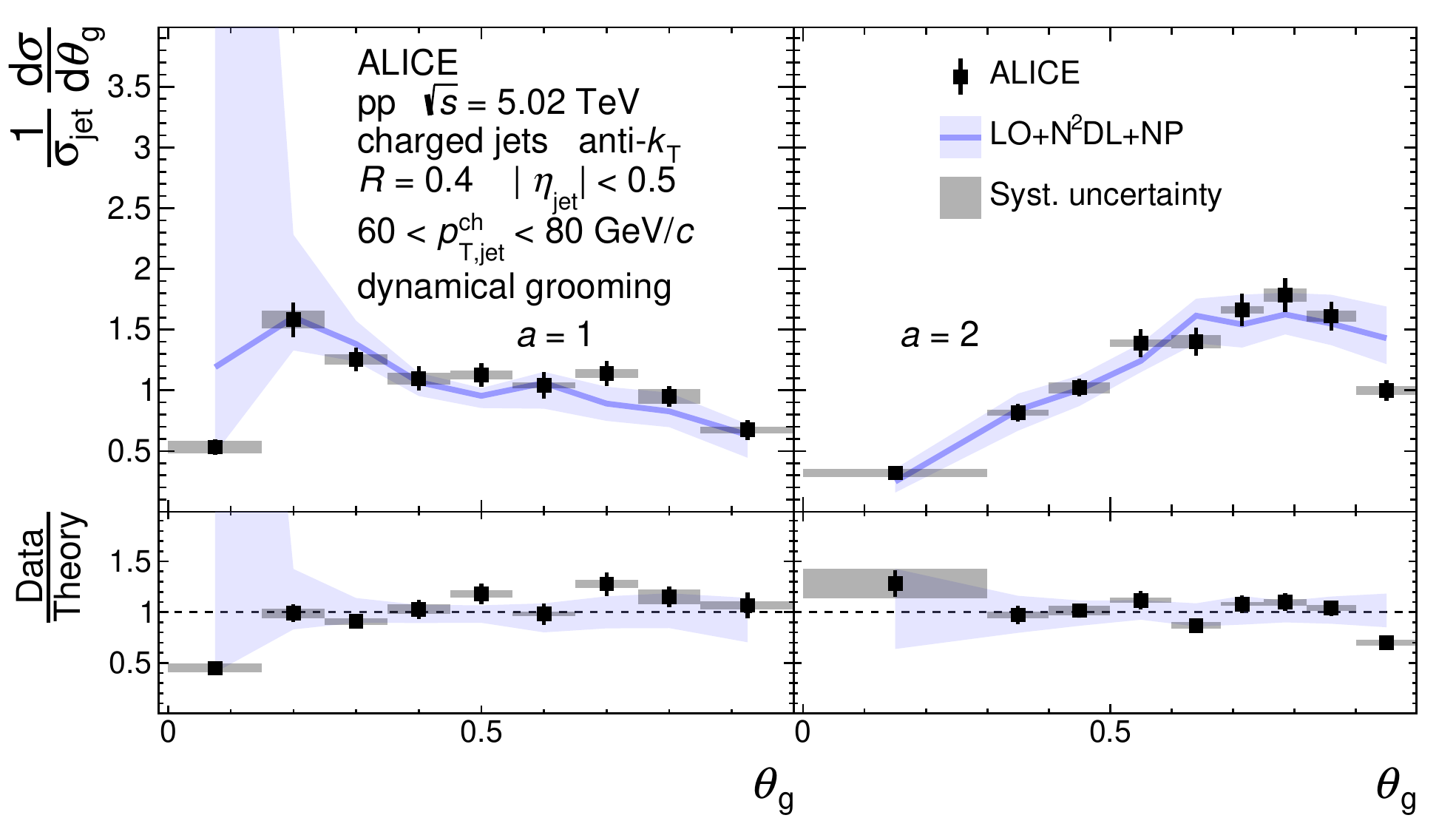}
\caption{ALICE measurements of \tg{} distributions 
in \pp{} collisions  at $\sqrt{s}=5.02\;$ TeV with dynamical grooming 
for two values of the
grooming parameter $a$, compared with pQCD calculations~\cite{Mehtar-Tani:2019rrk, Caucal:2021bae}.}
\label{fig:dg-tg-theory}
\end{figure}

In Figs.~\ref{fig:dg-zg-theory} and~\ref{fig:dg-tg-theory}, 
we compare the \zg{} and \tg{} distributions, respectively,
to pQCD calculations described in Ref.~\cite{Caucal:2021bae}.
The theoretical calculations include non-perturbative corrections
based on MC event generators, 
which are implemented in Ref.~\cite{Caucal:2021bae}.
The theoretical uncertainty bands account for scale variations together with non-perturbative effects, the latter generally being the dominant contribution.
The calculations generally describe the data within the precision of
the statistical and systematic uncertainties of the data
and the theoretical uncertainties of the calculation,
demonstrating that pQCD predictions, when coupled with corrections for non-perturbative effects, provide a sufficient description of the data
even at the moderate \pTchjet{} considered here.

\section{Conclusions}

We have presented new measurements of the groomed jet radius and
momentum splitting fraction in \pp{} collisions at {$\sqrt{s}=5.02$ TeV} with the ALICE detector at the Large Hadron Collider. 
We studied two grooming algorithms, soft drop and dynamical grooming, 
each with a variety of grooming settings in order to study their impact 
on soft- and wide-angle radiation.
These studies have provided the first measurement of a jet substructure
observable with the dynamical grooming procedure.
We compared these results to perturbative calculations that include
resummation of large logarithms at all orders in the strong coupling constant,
and generally found agreement of the theoretical predictions with the data 
in the perturbative regime.
This conclusion holds for all grooming settings considered.
However, the soft drop \tg{} distributions
increasingly deviate from the perturbative calculations at small \tg{}
as the grooming parameter $\beta$ is increased (corresponding to grooming away less collinear radiation).
This is in accordance with the predicted limitation of the perturbative calculation in describing the non-perturbative region,
and provides guidance for the regimes within which perturbative QCD
can be used to describe the observables.
These measurements can be used both to test future perturbative calculations and
models of non-perturbative effects, 
and can serve as a baseline reference for future measurements in heavy-ion collisions.

\newenvironment{acknowledgement}{\relax}{\relax}
\begin{acknowledgement}
\section*{Acknowledgements}
We gratefully acknowledge Paul Caucal, Kyle Lee, Felix Ringer, and Alba Soto-Ontoso, and Adam Takacs for providing theoretical predictions.

The ALICE Collaboration would like to thank all its engineers and technicians for their invaluable contributions to the construction of the experiment and the CERN accelerator teams for the outstanding performance of the LHC complex.
The ALICE Collaboration gratefully acknowledges the resources and support provided by all Grid centres and the Worldwide LHC Computing Grid (WLCG) collaboration.
The ALICE Collaboration acknowledges the following funding agencies for their support in building and running the ALICE detector:
A. I. Alikhanyan National Science Laboratory (Yerevan Physics Institute) Foundation (ANSL), State Committee of Science and World Federation of Scientists (WFS), Armenia;
Austrian Academy of Sciences, Austrian Science Fund (FWF): [M 2467-N36] and Nationalstiftung f\"{u}r Forschung, Technologie und Entwicklung, Austria;
Ministry of Communications and High Technologies, National Nuclear Research Center, Azerbaijan;
Conselho Nacional de Desenvolvimento Cient\'{\i}fico e Tecnol\'{o}gico (CNPq), Financiadora de Estudos e Projetos (Finep), Funda\c{c}\~{a}o de Amparo \`{a} Pesquisa do Estado de S\~{a}o Paulo (FAPESP) and Universidade Federal do Rio Grande do Sul (UFRGS), Brazil;
Bulgarian Ministry of Education and Science, within the National Roadmap for Research Infrastructures 2020¿2027 (object CERN), Bulgaria;
Ministry of Education of China (MOEC) , Ministry of Science \& Technology of China (MSTC) and National Natural Science Foundation of China (NSFC), China;
Ministry of Science and Education and Croatian Science Foundation, Croatia;
Centro de Aplicaciones Tecnol\'{o}gicas y Desarrollo Nuclear (CEADEN), Cubaenerg\'{\i}a, Cuba;
Ministry of Education, Youth and Sports of the Czech Republic, Czech Republic;
The Danish Council for Independent Research | Natural Sciences, the VILLUM FONDEN and Danish National Research Foundation (DNRF), Denmark;
Helsinki Institute of Physics (HIP), Finland;
Commissariat \`{a} l'Energie Atomique (CEA) and Institut National de Physique Nucl\'{e}aire et de Physique des Particules (IN2P3) and Centre National de la Recherche Scientifique (CNRS), France;
Bundesministerium f\"{u}r Bildung und Forschung (BMBF) and GSI Helmholtzzentrum f\"{u}r Schwerionenforschung GmbH, Germany;
General Secretariat for Research and Technology, Ministry of Education, Research and Religions, Greece;
National Research, Development and Innovation Office, Hungary;
Department of Atomic Energy Government of India (DAE), Department of Science and Technology, Government of India (DST), University Grants Commission, Government of India (UGC) and Council of Scientific and Industrial Research (CSIR), India;
National Research and Innovation Agency - BRIN, Indonesia;
Istituto Nazionale di Fisica Nucleare (INFN), Italy;
Japanese Ministry of Education, Culture, Sports, Science and Technology (MEXT) and Japan Society for the Promotion of Science (JSPS) KAKENHI, Japan;
Consejo Nacional de Ciencia (CONACYT) y Tecnolog\'{i}a, through Fondo de Cooperaci\'{o}n Internacional en Ciencia y Tecnolog\'{i}a (FONCICYT) and Direcci\'{o}n General de Asuntos del Personal Academico (DGAPA), Mexico;
Nederlandse Organisatie voor Wetenschappelijk Onderzoek (NWO), Netherlands;
The Research Council of Norway, Norway;
Commission on Science and Technology for Sustainable Development in the South (COMSATS), Pakistan;
Pontificia Universidad Cat\'{o}lica del Per\'{u}, Peru;
Ministry of Education and Science, National Science Centre and WUT ID-UB, Poland;
Korea Institute of Science and Technology Information and National Research Foundation of Korea (NRF), Republic of Korea;
Ministry of Education and Scientific Research, Institute of Atomic Physics, Ministry of Research and Innovation and Institute of Atomic Physics and University Politehnica of Bucharest, Romania;
Ministry of Education, Science, Research and Sport of the Slovak Republic, Slovakia;
National Research Foundation of South Africa, South Africa;
Swedish Research Council (VR) and Knut \& Alice Wallenberg Foundation (KAW), Sweden;
European Organization for Nuclear Research, Switzerland;
Suranaree University of Technology (SUT), National Science and Technology Development Agency (NSTDA), Thailand Science Research and Innovation (TSRI) and National Science, Research and Innovation Fund (NSRF), Thailand;
Turkish Energy, Nuclear and Mineral Research Agency (TENMAK), Turkey;
National Academy of  Sciences of Ukraine, Ukraine;
Science and Technology Facilities Council (STFC), United Kingdom;
National Science Foundation of the United States of America (NSF) and United States Department of Energy, Office of Nuclear Physics (DOE NP), United States of America.
In addition, individual groups or members have received support from:
Marie Sk\l{}odowska Curie, Strong 2020 - Horizon 2020, European Research Council (grant nos. 824093, 896850, 950692), European Union;
Academy of Finland (Center of Excellence in Quark Matter) (grant nos. 346327, 346328), Finland;
Programa de Apoyos para la Superaci\'{o}n del Personal Acad\'{e}mico, UNAM, Mexico.
\end{acknowledgement}

\bibliographystyle{utphys}
\bibliography{main.bib}

\clearpage
\newpage
\appendix

%
%

\section{The ALICE Collaboration}
\label{app:collab}
\begin{flushleft} 
\small

S.~Acharya\,\orcidlink{0000-0002-9213-5329}\,$^{\rm 124,131}$, 
D.~Adamov\'{a}\,\orcidlink{0000-0002-0504-7428}\,$^{\rm 86}$, 
A.~Adler$^{\rm 69}$, 
G.~Aglieri Rinella\,\orcidlink{0000-0002-9611-3696}\,$^{\rm 32}$, 
M.~Agnello\,\orcidlink{0000-0002-0760-5075}\,$^{\rm 29}$, 
N.~Agrawal\,\orcidlink{0000-0003-0348-9836}\,$^{\rm 50}$, 
Z.~Ahammed\,\orcidlink{0000-0001-5241-7412}\,$^{\rm 131}$, 
S.~Ahmad\,\orcidlink{0000-0003-0497-5705}\,$^{\rm 15}$, 
S.U.~Ahn\,\orcidlink{0000-0001-8847-489X}\,$^{\rm 70}$, 
I.~Ahuja\,\orcidlink{0000-0002-4417-1392}\,$^{\rm 37}$
% ...

\section*{Affiliation Notes}

$^{\rm I}$ Deceased\\
$^{\rm II}$ Also at: Max-Planck-Institut f\"{u}r Physik, Munich, Germany\\
$^{\rm III}$ Also at: Italian National Agency for New Technologies, Energy and Sustainable Economic Development (ENEA), Bologna, Italy\\
$^{\rm IV}$ Also at: Dipartimento DET del Politecnico di Torino, Turin, Italy\\
$^{\rm V}$ Also at: Department of Applied Physics, Aligarh Muslim University, Aligarh, India\\
$^{\rm VI}$ Also at: Institute of Theoretical Physics, University of Wroclaw, Poland\\
$^{\rm VII}$ Also at: An institution covered by a cooperation agreement with CERN\\

\section*{Collaboration Institutes}

$^{1}$ A.I. Alikhanyan National Science Laboratory (Yerevan Physics Institute) Foundation, Yerevan, Armenia\\
$^{2}$ AGH University of Science and Technology, Cracow, Poland\\
$^{3}$ Bogolyubov Institute for Theoretical Physics, National Academy of Sciences of Ukraine, Kiev, Ukraine\\
$^{4}$ Bose Institute, Department of Physics  and Centre for Astroparticle Physics and Space Science (CAPSS), Kolkata, India\\
$^{5}$ California Polytechnic State University, San Luis Obispo, California, United States\\
$^{6}$ Central China Normal University, Wuhan, China\\
$^{7}$ Centro de Aplicaciones Tecnol\'{o}gicas y Desarrollo Nuclear (CEADEN), Havana, Cuba\\
$^{8}$ Centro de Investigaci\'{o}n y de Estudios Avanzados (CINVESTAV), Mexico City and M\'{e}rida, Mexico\\
$^{9}$ Chicago State University, Chicago, Illinois, United States\\
$^{10}$ China Institute of Atomic Energy, Beijing, China\\
$^{11}$ Chungbuk National University, Cheongju, Republic of Korea\\
$^{12}$ Comenius University Bratislava, Faculty of Mathematics, Physics and Informatics, Bratislava, Slovak Republic\\
$^{13}$ COMSATS University Islamabad, Islamabad, Pakistan\\
$^{14}$ Creighton University, Omaha, Nebraska, United States\\
$^{15}$ Department of Physics, Aligarh Muslim University, Aligarh, India\\
$^{16}$ Department of Physics, Pusan National University, Pusan, Republic of Korea\\
$^{17}$ Department of Physics, Sejong University, Seoul, Republic of Korea\\
$^{18}$ Department of Physics, University of California, Berkeley, California, United States\\
$^{19}$ Department of Physics, University of Oslo, Oslo, Norway\\
$^{20}$ Department of Physics and Technology, University of Bergen, Bergen, Norway\\
$^{21}$ Dipartimento di Fisica, Universit\`{a} di Pavia, Pavia, Italy\\
$^{22}$ Dipartimento di Fisica dell'Universit\`{a} and Sezione INFN, Cagliari, Italy\\
$^{23}$ Dipartimento di Fisica dell'Universit\`{a} and Sezione INFN, Trieste, Italy\\
$^{24}$ Dipartimento di Fisica dell'Universit\`{a} and Sezione INFN, Turin, Italy\\
$^{25}$ Dipartimento di Fisica e Astronomia dell'Universit\`{a} and Sezione INFN, Bologna, Italy\\
$^{26}$ Dipartimento di Fisica e Astronomia dell'Universit\`{a} and Sezione INFN, Catania, Italy\\
$^{27}$ Dipartimento di Fisica e Astronomia dell'Universit\`{a} and Sezione INFN, Padova, Italy\\
$^{28}$ Dipartimento di Fisica `E.R.~Caianiello' dell'Universit\`{a} and Gruppo Collegato INFN, Salerno, Italy\\
$^{29}$ Dipartimento DISAT del Politecnico and Sezione INFN, Turin, Italy\\
$^{30}$ Dipartimento di Scienze MIFT, Universit\`{a} di Messina, Messina, Italy\\
$^{31}$ Dipartimento Interateneo di Fisica `M.~Merlin' and Sezione INFN, Bari, Italy\\
$^{32}$ European Organization for Nuclear Research (CERN), Geneva, Switzerland\\
$^{33}$ Faculty of Electrical Engineering, Mechanical Engineering and Naval Architecture, University of Split, Split, Croatia\\
$^{34}$ Faculty of Engineering and Science, Western Norway University of Applied Sciences, Bergen, Norway\\
$^{35}$ Faculty of Nuclear Sciences and Physical Engineering, Czech Technical University in Prague, Prague, Czech Republic\\
$^{36}$ Faculty of Physics, Sofia University, Sofia, Bulgaria\\
$^{37}$ Faculty of Science, P.J.~\v{S}af\'{a}rik University, Ko\v{s}ice, Slovak Republic\\
$^{38}$ Frankfurt Institute for Advanced Studies, Johann Wolfgang Goethe-Universit\"{a}t Frankfurt, Frankfurt, Germany\\
$^{39}$ Fudan University, Shanghai, China\\
$^{40}$ Gangneung-Wonju National University, Gangneung, Republic of Korea\\
$^{41}$ Gauhati University, Department of Physics, Guwahati, India\\
$^{42}$ Helmholtz-Institut f\"{u}r Strahlen- und Kernphysik, Rheinische Friedrich-Wilhelms-Universit\"{a}t Bonn, Bonn, Germany\\
$^{43}$ Helsinki Institute of Physics (HIP), Helsinki, Finland\\
$^{44}$ High Energy Physics Group,  Universidad Aut\'{o}noma de Puebla, Puebla, Mexico\\
$^{45}$ Horia Hulubei National Institute of Physics and Nuclear Engineering, Bucharest, Romania\\
$^{46}$ Indian Institute of Technology Bombay (IIT), Mumbai, India\\
$^{47}$ Indian Institute of Technology Indore, Indore, India\\
$^{48}$ INFN, Laboratori Nazionali di Frascati, Frascati, Italy\\
$^{49}$ INFN, Sezione di Bari, Bari, Italy\\
$^{50}$ INFN, Sezione di Bologna, Bologna, Italy\\
$^{51}$ INFN, Sezione di Cagliari, Cagliari, Italy\\
$^{52}$ INFN, Sezione di Catania, Catania, Italy\\
$^{53}$ INFN, Sezione di Padova, Padova, Italy\\
$^{54}$ INFN, Sezione di Pavia, Pavia, Italy\\
$^{55}$ INFN, Sezione di Torino, Turin, Italy\\
$^{56}$ INFN, Sezione di Trieste, Trieste, Italy\\
$^{57}$ Inha University, Incheon, Republic of Korea\\
$^{58}$ Institute for Gravitational and Subatomic Physics (GRASP), Utrecht University/Nikhef, Utrecht, Netherlands\\
$^{59}$ Institute of Experimental Physics, Slovak Academy of Sciences, Ko\v{s}ice, Slovak Republic\\
$^{60}$ Institute of Physics, Homi Bhabha National Institute, Bhubaneswar, India\\
$^{61}$ Institute of Physics of the Czech Academy of Sciences, Prague, Czech Republic\\
$^{62}$ Institute of Space Science (ISS), Bucharest, Romania\\
$^{63}$ Institut f\"{u}r Kernphysik, Johann Wolfgang Goethe-Universit\"{a}t Frankfurt, Frankfurt, Germany\\
$^{64}$ Instituto de Ciencias Nucleares, Universidad Nacional Aut\'{o}noma de M\'{e}xico, Mexico City, Mexico\\
$^{65}$ Instituto de F\'{i}sica, Universidade Federal do Rio Grande do Sul (UFRGS), Porto Alegre, Brazil\\
$^{66}$ Instituto de F\'{\i}sica, Universidad Nacional Aut\'{o}noma de M\'{e}xico, Mexico City, Mexico\\
$^{67}$ iThemba LABS, National Research Foundation, Somerset West, South Africa\\
$^{68}$ Jeonbuk National University, Jeonju, Republic of Korea\\
$^{69}$ Johann-Wolfgang-Goethe Universit\"{a}t Frankfurt Institut f\"{u}r Informatik, Fachbereich Informatik und Mathematik, Frankfurt, Germany\\
$^{70}$ Korea Institute of Science and Technology Information, Daejeon, Republic of Korea\\
$^{71}$ KTO Karatay University, Konya, Turkey\\
$^{72}$ Laboratoire de Physique des 2 Infinis, Ir\`{e}ne Joliot-Curie, Orsay, France\\
$^{73}$ Laboratoire de Physique Subatomique et de Cosmologie, Universit\'{e} Grenoble-Alpes, CNRS-IN2P3, Grenoble, France\\
$^{74}$ Lawrence Berkeley National Laboratory, Berkeley, California, United States\\
$^{75}$ Lund University Department of Physics, Division of Particle Physics, Lund, Sweden\\
$^{76}$ Nagasaki Institute of Applied Science, Nagasaki, Japan\\
$^{77}$ Nara Women{'}s University (NWU), Nara, Japan\\
$^{78}$ National and Kapodistrian University of Athens, School of Science, Department of Physics , Athens, Greece\\
$^{79}$ National Centre for Nuclear Research, Warsaw, Poland\\
$^{80}$ National Institute of Science Education and Research, Homi Bhabha National Institute, Jatni, India\\
$^{81}$ National Nuclear Research Center, Baku, Azerbaijan\\
$^{82}$ National Research and Innovation Agency - BRIN, Jakarta, Indonesia\\
$^{83}$ Niels Bohr Institute, University of Copenhagen, Copenhagen, Denmark\\
$^{84}$ Nikhef, National institute for subatomic physics, Amsterdam, Netherlands\\
$^{85}$ Nuclear Physics Group, STFC Daresbury Laboratory, Daresbury, United Kingdom\\
$^{86}$ Nuclear Physics Institute of the Czech Academy of Sciences, Husinec-\v{R}e\v{z}, Czech Republic\\
$^{87}$ Oak Ridge National Laboratory, Oak Ridge, Tennessee, United States\\
$^{88}$ Ohio State University, Columbus, Ohio, United States\\
$^{89}$ Physics department, Faculty of science, University of Zagreb, Zagreb, Croatia\\
$^{90}$ Physics Department, Panjab University, Chandigarh, India\\
$^{91}$ Physics Department, University of Jammu, Jammu, India\\
$^{92}$ Physics Department, University of Rajasthan, Jaipur, India\\
$^{93}$ Physics Program and International Institute for Sustainability with Knotted Chiral Meta Matter (SKCM2), Hiroshima University, Hiroshima, Japan\\
$^{94}$ Physikalisches Institut, Eberhard-Karls-Universit\"{a}t T\"{u}bingen, T\"{u}bingen, Germany\\
$^{95}$ Physikalisches Institut, Ruprecht-Karls-Universit\"{a}t Heidelberg, Heidelberg, Germany\\
$^{96}$ Physik Department, Technische Universit\"{a}t M\"{u}nchen, Munich, Germany\\
$^{97}$ Politecnico di Bari and Sezione INFN, Bari, Italy\\
$^{98}$ Research Division and ExtreMe Matter Institute EMMI, GSI Helmholtzzentrum f\"ur Schwerionenforschung GmbH, Darmstadt, Germany\\
$^{99}$ Saha Institute of Nuclear Physics, Homi Bhabha National Institute, Kolkata, India\\
$^{100}$ School of Physics and Astronomy, University of Birmingham, Birmingham, United Kingdom\\
$^{101}$ Secci\'{o}n F\'{\i}sica, Departamento de Ciencias, Pontificia Universidad Cat\'{o}lica del Per\'{u}, Lima, Peru\\
$^{102}$ Stefan Meyer Institut f\"{u}r Subatomare Physik (SMI), Vienna, Austria\\
$^{103}$ SUBATECH, IMT Atlantique, Nantes Universit\'{e}, CNRS-IN2P3, Nantes, France\\
$^{104}$ Suranaree University of Technology, Nakhon Ratchasima, Thailand\\
$^{105}$ Technical University of Ko\v{s}ice, Ko\v{s}ice, Slovak Republic\\
$^{106}$ The Henryk Niewodniczanski Institute of Nuclear Physics, Polish Academy of Sciences, Cracow, Poland\\
$^{107}$ The University of Texas at Austin, Austin, Texas, United States\\
$^{108}$ Universidad Aut\'{o}noma de Sinaloa, Culiac\'{a}n, Mexico\\
$^{109}$ Universidade de S\~{a}o Paulo (USP), S\~{a}o Paulo, Brazil\\
$^{110}$ Universidade Estadual de Campinas (UNICAMP), Campinas, Brazil\\
$^{111}$ Universidade Federal do ABC, Santo Andre, Brazil\\
$^{112}$ University of Cape Town, Cape Town, South Africa\\
$^{113}$ University of Houston, Houston, Texas, United States\\
$^{114}$ University of Jyv\"{a}skyl\"{a}, Jyv\"{a}skyl\"{a}, Finland\\
$^{115}$ University of Kansas, Lawrence, Kansas, United States\\
$^{116}$ University of Liverpool, Liverpool, United Kingdom\\
$^{117}$ University of Science and Technology of China, Hefei, China\\
$^{118}$ University of South-Eastern Norway, Kongsberg, Norway\\
$^{119}$ University of Tennessee, Knoxville, Tennessee, United States\\
$^{120}$ University of the Witwatersrand, Johannesburg, South Africa\\
$^{121}$ University of Tokyo, Tokyo, Japan\\
$^{122}$ University of Tsukuba, Tsukuba, Japan\\
$^{123}$ University Politehnica of Bucharest, Bucharest, Romania\\
$^{124}$ Universit\'{e} Clermont Auvergne, CNRS/IN2P3, LPC, Clermont-Ferrand, France\\
$^{125}$ Universit\'{e} de Lyon, CNRS/IN2P3, Institut de Physique des 2 Infinis de Lyon, Lyon, France\\
$^{126}$ Universit\'{e} de Strasbourg, CNRS, IPHC UMR 7178, F-67000 Strasbourg, France, Strasbourg, France\\
$^{127}$ Universit\'{e} Paris-Saclay Centre d'Etudes de Saclay (CEA), IRFU, D\'{e}partment de Physique Nucl\'{e}aire (DPhN), Saclay, France\\
$^{128}$ Universit\`{a} degli Studi di Foggia, Foggia, Italy\\
$^{129}$ Universit\`{a} del Piemonte Orientale, Vercelli, Italy\\
$^{130}$ Universit\`{a} di Brescia, Brescia, Italy\\
$^{131}$ Variable Energy Cyclotron Centre, Homi Bhabha National Institute, Kolkata, India\\
$^{132}$ Warsaw University of Technology, Warsaw, Poland\\
$^{133}$ Wayne State University, Detroit, Michigan, United States\\
$^{134}$ Westf\"{a}lische Wilhelms-Universit\"{a}t M\"{u}nster, Institut f\"{u}r Kernphysik, M\"{u}nster, Germany\\
$^{135}$ Wigner Research Centre for Physics, Budapest, Hungary\\
$^{136}$ Yale University, New Haven, Connecticut, United States\\
$^{137}$ Yonsei University, Seoul, Republic of Korea\\
$^{138}$  Zentrum  f\"{u}r Technologie und Transfer (ZTT), Worms, Germany\\
$^{139}$ Affiliated with an institute covered by a cooperation agreement with CERN\\
$^{140}$ Affiliated with an international laboratory covered by a cooperation agreement with CERN.\\

\end{flushleft} 

\end{document}